\documentclass{article}

\usepackage[nonatbib,nonanonymous]{paper}
\makeatletter\renewcommand{\@noticestring}{}\makeatother
\nolinenumbers

\usepackage[utf8]{inputenc}
\usepackage[T1]{fontenc}
\usepackage{hyperref}
\usepackage{url}
\usepackage{booktabs}
\usepackage{amsfonts}
\usepackage{amsmath}
\usepackage{nicefrac}
\usepackage{microtype}
\usepackage[table]{xcolor}
\usepackage{graphicx}
\usepackage{subcaption}
\usepackage{algorithm}
\usepackage{algorithmic}
\usepackage{amssymb}
\usepackage{pifont}
\usepackage{colortbl}
\usepackage{graphicx}
\usepackage{caption}

\usepackage{cleveref}
\usepackage[numbers]{natbib}

\newcommand{\method}{\textsc{ScenA}}

\definecolor{betacol}{HTML}{1F77B4}
\definecolor{rightcol}{HTML}{2CA02C}
\definecolor{leftcol}{HTML}{FF7F0E}
\definecolor{leftmostcol}{HTML}{D62728}

\title{Reference-Driven Multi-Speaker Audio Scene Generation from In-the-Wild Priors}

\author{
  \bfseries
  Michael Finkelson$^{1,2}$\enspace
  Daniel Segal$^{1}$\enspace
  Eitan Richardson$^{1}$\enspace
  Shahar Armon$^{1}$\enspace
  Nani Goldring$^{1}$\\[3pt]
  \bfseries
  Poriya Panet$^{1}$\enspace
  Nir Zabari$^{1}$\enspace
  Benjamin Brazowski$^{1}$\enspace
  Or Patashnik$^{2}$\enspace
  Yoav HaCohen$^{1}$\\[5pt]
  \normalfont\normalsize $^{1}$Lightricks \qquad $^{2}$Tel Aviv University
}

\begin{document}

\maketitle

\begin{figure}[h]
    \centering
    \includegraphics[width=\linewidth]{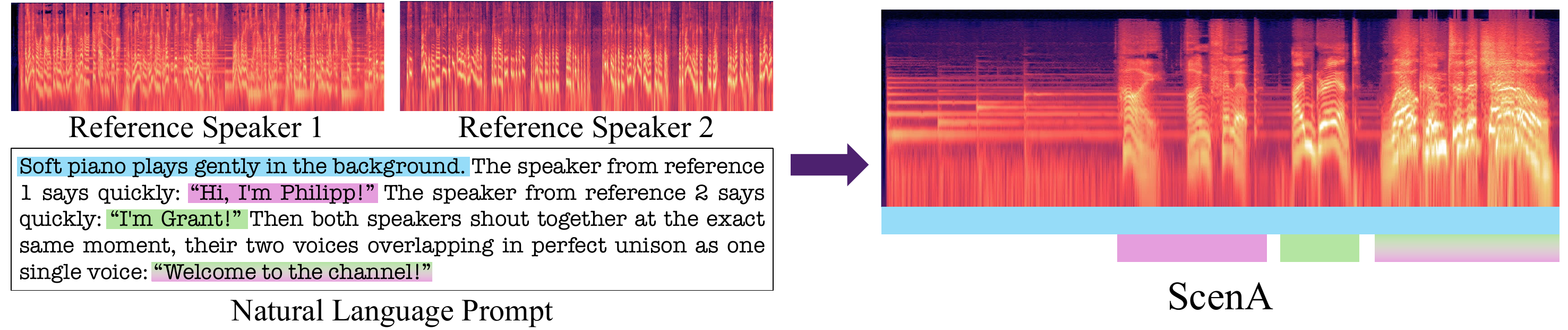}
    \caption{Our \method{} framework transforms free-form natural language prompts and a set of reference voices into rich, multi-speaker conversational scenes. The prompt alone determines which reference speaks where, with no per-turn tags, transcripts, or identity encoders. This natural language interface enables complex human interactions, including overlapping speech, spontaneous paralinguistic events, and scene-level ambient sound.}
    \label{fig:teaser}
\end{figure}

\begin{abstract}
Existing multi-speaker dialogue systems bind speakers to utterances through structured supervision: per-turn tags, multi-stream transcriptions, or learnable speaker embeddings. These systems operate within speech-only pipelines that produce clean vocal sequences without the ambient texture of real conversations. We take a different approach. Our method, \method{}, conditions a text-to-audio flow-matching foundation model, pretrained on large-scale in-the-wild data, directly on multiple reference voices and a free-form natural language prompt that describes an entire multi-speaker audio scene. Leveraging such a foundational model allows us to inherit its capacity for natural, non-studio audio: background noise, room acoustics, overlapping dialogue, and spontaneous paralinguistic events, while adding multi-speaker control without any per-turn structure. Concretely, reference latents are concatenated into the model's token sequence and distinguished by lightweight identity-aware positional encodings. However, we identify a critical obstacle to this approach: the \textit{Reference Shortcut}.
During training under standard noise schedules, the model can identify the matching reference by acoustic similarity to the noisy target, bypassing the text prompt entirely. We address this with a high-noise-biased timestep distribution that forces the model to rely on the text prompt for speaker assignment. We evaluate \method{} on the CoVoMix2-Dialogue benchmark, showing that it outperforms existing multi-speaker systems on speaker-binding metrics while generating rich conversational audio with overlapping speech, emotional vocalizations, and ambient sound. Our results demonstrate the advantage of using a general-purpose audio model conditioned on a free-form scene description, rather than passing structured dialog scripts through a speech-only pipeline.

\noindent\textbf{Project page:} \url{https://finmickey.github.io/scena/}
\end{abstract}

\section{Introduction}
\label{sec:intro}

A real multi-speaker conversation is more than ordered turns of speech: voices overlap, laughter cuts in, and the room itself shapes how it all sounds.
However, existing speech generation systems capture little of this.
Zero-shot voice cloning models address only the single-speaker case~\cite{wang2023valle,du2024cosyvoice2,chen2024f5tts,casanova2024xtts}, leaving multi-speaker conversations to be assembled segment-by-segment outside the model.
Recent dialogue-TTS systems generate multi-turn conversations directly~\cite{zhang2026mossttsd,peng2025vibevoice,zhu2025zipvoicedialog,zhang2025covomix2}, but bind speakers to utterances through \emph{structured} supervision (per-turn speaker tags, multi-stream transcriptions, or learnable speaker-turn embeddings).
All are speech-only pipelines, producing clean vocal tracks stripped of the ambient texture real conversations carry.

We take a different starting point with \method{}, our flow-matching framework for multi-speaker audio scene generation.
Flow-matching~\cite{lipman2023flow,liu2023flow} text-to-audio foundation models~\cite{hacohen2026ltx2}, pretrained on large-scale in-the-wild audio, already capture the texture of natural sound scenes.
What they lack is a way to tie specific voices to specific roles.
\method{} adds this capability with a deliberately minimal interface: reference latents are concatenated with the input latents and distinguished by lightweight identity-aware positional encodings.
A single free-form natural language prompt describes the entire scene: who speaks, what is said, and what else is in the room.
No per-turn tags, no multi-stream transcripts, and no identity encoders or reference-side adapters~\cite{ye2023ipadapter,wang2024msdiffusion}; the prompt alone determines which reference speaks where (\Cref{fig:teaser}).

Our experiments show that na\"ively training this design fails to learn which reference speaks where, due to a previously unrecognized failure mode of reference-conditioned flow matching that we call the \emph{reference shortcut}.
We find that under the standard logit-normal timestep distribution~\cite{esser2024scaling}, the noised target retains enough acoustic information for the model to pick the matching reference by similarity.
This shortcut sidesteps the text prompt entirely, and
yields low training loss but catastrophic inference.
At test time, generation begins from pure noise, where the shortcut is unavailable and text is the only signal that can resolve which voice goes where.
Text, however, is precisely the signal the model has learned to ignore.
To locate the noise levels where the shortcut works, we train a small probe on frozen audio features and find that it can match references to targets by similarity across the entire low-to-moderate noise range — exactly where standard training concentrates. A noise-schedule ablation then shows that binding-aware metrics improve monotonically as we shift training mass toward higher noise.
We close this shortcut with a \emph{high-noise-biased} timestep distribution, a Beta+Uniform mixture that concentrates training on noise levels where the target is uninformative and text is the only binding signal.

We evaluate our method on the public \textsc{CoVoMix2-Dialogue} benchmark, where \method{} matches or surpasses current multi-speaker dialog baselines on every speaker-binding metric.
The advantage widens on a harder in-the-wild reference subset, where studio-clean references give way to noisy real-world recordings.
Beyond binding, \method{} generates overlapping dialogue, spontaneous paralinguistic events (laughter, sighs, breaths), and scene-level ambient sound jointly with the conversation.
We encourage readers to visit our project page for the full spectrum of capabilities beyond two-speaker dialogue.

\enlargethispage{2\baselineskip}
\section{Related Work}
\label{sec:related}

\paragraph{Multi-speaker speech generation.}
Zero-shot voice cloning has converged on two architectural families.
Autoregressive zero-shot TTS systems (VALL-E~\cite{wang2023valle,chen2024valle2}, Seed-TTS~\cite{anastassiou2024seedtts}, Spark-TTS~\cite{wang2025sparktts}, CosyVoice~2/3~\cite{du2024cosyvoice2,du2025cosyvoice3}, VoiceStar~\cite{peng2025voicestar}, XTTS~\cite{casanova2024xtts}, MiniMax-Speech~\cite{zhang2025minimaxspeech}) autoregressively generate discrete speech tokens from a single speaker reference clip.
Flow-matching and diffusion generators (Voicebox~\cite{le2023voicebox}, E2-TTS~\cite{eskimez2024e2tts}, F5-TTS~\cite{chen2024f5tts}, NaturalSpeech~3~\cite{ju2024naturalspeech3}, MegaTTS~3~\cite{jiang2025megatts3}, ZipVoice~\cite{zhu2025zipvoice}, StyleTTS~2~\cite{li2023styletts2}) produce mel-spectrograms or latents in a single non-autoregressive pass.
Both families condition on a \emph{single} reference; multi-speaker conversations are obtained post hoc by synthesizing each speaker's segments independently and concatenating, an arrangement that is incompatible with overlapping speech, shared acoustic environments, and scene-level descriptions.
A more recent wave of dialog-TTS systems generates multi-turn conversations directly~\cite{ju2025mooncast,zhang2024covomix,zhang2025covomix2,zhu2025zipvoicedialog,peng2025vibevoice,xie2025fireredtts2,xie2025soulxpodcast,xie2025dialospeech,zhang2026mossttsd,yu2025joyvoice,narilabs2024dia}, but binds speakers to utterances through \emph{structured} supervision: per-turn speaker tags (e.g., \texttt{[S1]}/\texttt{[S2]}), multi-stream transcriptions, learnable speaker-turn embeddings, or LLM-generated annotations.

\newpage
\paragraph{Reference-conditioned generation.}
Beyond zero-shot TTS, reference conditioning has been studied across audio generation more broadly.
Audiobox~\cite{vyas2023audiobox} is closest to our setting: a flow-matching audio model that conditions on a text caption together with a single voice prompt to jointly synthesise speech, sound, and music.
MusicGen~\cite{copet2023musicgen} pairs text with a melody reference for music generation; AudioLDM~2~\cite{liu2023audioldm2} conditions sound generation on text together with an audio prompt.
All of these condition on a \emph{single} reference.
We are, to our knowledge, the first to address the multi-reference audio setting where natural language alone determines speaker assignment.

Parallel work in the image and video domains is informative for our design choices.
IP-Adapter~\cite{ye2023ipadapter} introduces decoupled cross-attention dedicated to image references; InstantID~\cite{wang2024instantid} and PhotoMaker~\cite{li2023photomaker} use face-specific encoders for identity preservation.
Multi-subject methods rely on additional structure beyond text: bounding-box layouts in MS-Diffusion~\cite{wang2024msdiffusion}, segmentation maps in MuDI~\cite{jang2024mudi}, localized cross-attention at training and image-augmented prompt tokens at inference in FastComposer~\cite{xiao2023fastcomposer}, and region-aware masked guidance for video in MAGREF~\cite{deng2025magref}.
Closer in spirit to our setup, three image-domain methods condition transformers on references in context. 
OmniGen~\cite{xiao2024omnigen} replaces each \texttt{<|image\_k|>} marker in the prompt with that reference image's tokens, producing a single sequence that interleaves text and image references.
UNO~\cite{wu2025uno} concatenates references in the attention sequence with offset rotary positions, and In-Context LoRA~\cite{huang2024incontextlora} stitches references and target into a composite image; both bind references by matching descriptive subject names in the prompt (\emph{the toy}, \emph{the man with blond hair}) to the reference list. This requires references to be separately describable in language, which fails when they share content, as is typical in a multi-speaker dialog.

\method{} binds references through indexed mentions (\emph{reference 1}, \emph{reference 2}) that need not describe their content, and ties them to dynamic spans of a jointly generated multi-speaker output.
This setting exposes a flow-matching-specific failure mode, the reference shortcut (\S\ref{sec:shortcut}), in which the model can bypass the prompt by matching references against the noised target. We close it
by modifying the timestep distribution alone, rather than by adding an explicit binding mechanism.

\paragraph{Timestep distributions and noise schedules.}
The choice of noise schedule has a large effect on diffusion-model training.
EDM~\cite{karras2022elucidating} proposes log-normal sigma sampling for image generation; Stable Diffusion~3~\cite{esser2024scaling} introduces the logit-normal timestep distribution for flow matching, with a mild upward shift at higher resolutions; Min-SNR weighting~\cite{hang2023efficient} reweights the loss by signal-to-noise ratio; and Simple Diffusion~\cite{hoogeboom2023simple} shifts schedules toward higher noise for high-resolution images.
All of these choices are motivated by \emph{generation quality} or \emph{training stability}.

\section{Method}
\label{sec:method}

\begin{figure}[!ht]
    \centering
    \includegraphics[width=\linewidth]{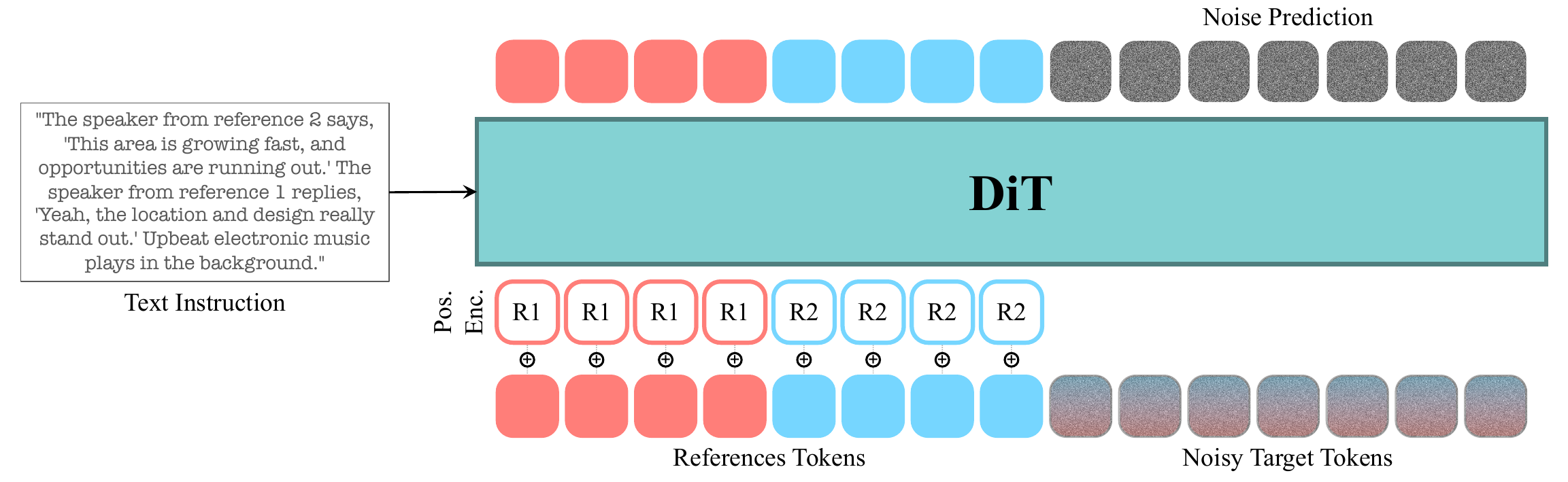}
    \caption{Our framework utilizes a DiT to synthesize reference-driven conversational scenes. Reference latents (red, blue) are concatenated with noisy target tokens, and identity-aware positional encodings ($R1, R2$) distinguish speakers. Text conditions the model via cross-attention, and training with a flow-matching loss enables transformation into high-fidelity conversational audio. }
    \label{fig:method}
\end{figure}

We describe our approach in four parts: the reference conditioning mechanism (\S\ref{sec:conditioning}), the reference shortcut and its analysis (\S\ref{sec:shortcut}), the timestep distribution that eliminates it (\S\ref{sec:recipe}), and the multi-reference dataset on which we train (\S\ref{sec:dataset}).
\Cref{fig:method} provides an overview.

\subsection{Reference Conditioning via Latent Concatenation}
\label{sec:conditioning}

\paragraph{Setup.}
We build on a pretrained audio diffusion transformer~\cite{hacohen2026ltx2} that operates on a sequence of latent tokens.
A target audio clip is encoded into a latent sequence $\mathbf{z}_0 \in \mathbb{R}^{N \times D}$ by the backbone's variational autoencoder (VAE), where $N$ is the number of tokens and $D$ is the latent dimension.
A text prompt $\mathbf{c}$ is encoded by a frozen text encoder and conditions generation through cross-attention; reference clips are encoded by the same VAE as the target.

\paragraph{Reference injection.}
Given $K \leq K_\text{max}$ reference speaker clips $\{r_1, \ldots, r_K\}$, each is encoded into a latent sequence $\mathbf{r}_k \in \mathbb{R}^{N_k \times D}$.
The full input to the transformer is the concatenation:
\begin{equation}
    \mathbf{z}_\text{input} = [\mathbf{z}_t;\; \mathbf{r}_1;\; \ldots;\; \mathbf{r}_K],
    \label{eq:concat}
\end{equation}
where $\mathbf{z}_t$ is the noised target at timestep $t$. Only the target is noised; the reference latents $\mathbf{r}_k$ are passed clean.
All tokens participate in the transformer's self-attention, allowing the model to attend freely between target and reference tokens.

\paragraph{Positional encoding.}
To distinguish the references from one another and from the target, we associate each reference slot $k$ with a learned embedding $e_k \in \mathbb{R}^{D}$, broadcast across the $N_k$ tokens of $\mathbf{r}_k$ and added immediately after the linear projection into the transformer hidden dimension:
\begin{equation}
    \mathbf{r}_k \leftarrow \mathbf{r}_k + e_k,
    \qquad k = 1, \ldots, K_\text{max},
    \label{eq:abs_emb}
\end{equation}
while the target $\mathbf{z}_t$ receives no additive embedding.
This adds only a negligible number of parameters and leaves the rest of the architecture unchanged.
We ablate the choice of slot encoding in Appendix~\ref{sec:exp_slot_encoding}, comparing this additive embedding against a RoPE-based alternative and a no-positional baseline.

\paragraph{Generation through natural language.}
A single text prompt drives the entire generation.
It describes the scene holistically (ambient sounds, speaker turns, content, and affect) and refers to each reference voice by an ordinary textual mention such as \emph{reference 1}.
For example:
\begin{quote}
\textit{``Ocean waves crash gently on the shore. Seagulls call in the distance. The speaker from reference 1 takes a deep breath and says: `This is exactly what I needed today.' The speaker from reference 2 hums in agreement: `Yeah, no emails out here.' Another wave rolls in slowly.''}
\end{quote}
This prompt, together with the concatenated reference latents (Eq.~\ref{eq:concat}) and their slot embeddings (Eq.~\ref{eq:abs_emb}), is the entire input to the model.
No special tokens, no identity-preserving adapters~\cite{ye2023ipadapter}, no per-segment transcripts, and no spatial or temporal supervision~\cite{wang2024msdiffusion, jang2024mudi} are used.
The model produces the entire scene in a single forward pass, including overlapping speech, natural turn-taking, and scene-level ambient sound.
Existing multi-speaker systems typically achieve these qualities only through external structure or post-hoc concatenation.
In \method{}, binding is delegated entirely to training, which, as we show next, requires careful design of the timestep distribution.

\subsection{The Reference Shortcut Challenge}
\label{sec:shortcut}

\paragraph{Flow matching background.}
In flow matching~\cite{lipman2023flow, liu2023flow}, training constructs a noised sample $\mathbf{z}_t = (1 - t)\mathbf{z}_0 + t\boldsymbol{\epsilon}$ at timestep $t \in [0, 1]$, where $\boldsymbol{\epsilon} \sim \mathcal{N}(0, \mathbf{I})$.
The model $f_\theta$ is trained to predict the velocity field $\mathbf{v} = \boldsymbol{\epsilon} - \mathbf{z}_0$ that transports noise to data.
The training objective is:
\begin{equation}
    \mathcal{L} = \mathbb{E}_{t \sim p(t),\, \mathbf{z}_0,\, \boldsymbol{\epsilon}} \left[ \| f_\theta(\mathbf{z}_t, t, \mathbf{c}) - \mathbf{v} \|^2 \right],
\end{equation}
where $p(t)$ is the timestep distribution.
Standard practice~\cite{esser2024scaling} uses a logit-normal distribution $t \sim \sigma(\mathcal{N}(\mu, s^2))$, often with a mild shift of $\mu$ toward higher noise.
In all such variants, the bulk of training mass remains concentrated on intermediate timesteps, where the denoising task is most informative for generation quality.
\vspace{20pt}
\paragraph{The shortcut mechanism.}

We now describe a shortcut available in our setup that, if not addressed, lets the model bypass the text prompt entirely.
Binding a reference through the prompt requires the model to compose information across two attention paths.
Through cross-attention, it must associate the phrase \emph{``reference $k$''} in the text with the slot embedding $e_k$ carried by the $k$-th reference's tokens.
Then, through self-attention, the target tokens must locate the reference bearing $e_k$ and route their queries toward it.
By contrast, the shortcut requires nothing more than what self-attention already does by default: when the noised target still carries acoustic traces of $\mathbf{z}_0$, its tokens are most similar to the
matching reference's, and self-attention routes them there in a single step, with no contribution from the text path.
Both routes minimize the training loss; the optimizer favors the simpler one whenever it is available.

Consider a training example with target $\mathbf{z}_0$, references $\{\mathbf{r}_1, \ldots, \mathbf{r}_K\}$, and text prompt $\mathbf{c}$ mentioning ``reference $k$''.
At timestep $t$, the noised target is $\mathbf{z}_t = (1-t)\mathbf{z}_0 + t\boldsymbol{\epsilon}$, interpolating between the clean target at $t = 0$ and pure noise at $t = 1$.
The shortcut is available whenever each speaker's segment of $\mathbf{z}_t$ remains discriminable from non-matching references; if this holds across most of $[0, 1]$, the prompt is rarely needed during training, and the model has little
reason to learn the two-path routing.
We quantify this with a probe that measures, at each $t$, whether $\mathbf{z}_t$ still carries enough of $\mathbf{z}_0$ to identify its source reference.

The probe is a binary classifier that mirrors the implicit denoising-time choice: given $\mathbf{z}_t$ and two candidate references (one matching, one from a different sample), predict the match.
We use the first 8 of the 48 transformer blocks of our backbone with a small classification head; full setup is in Appendix~\ref{sec:exp_probe}.
\begin{figure}[H]
    \centering
    \includegraphics[width=0.95\linewidth]{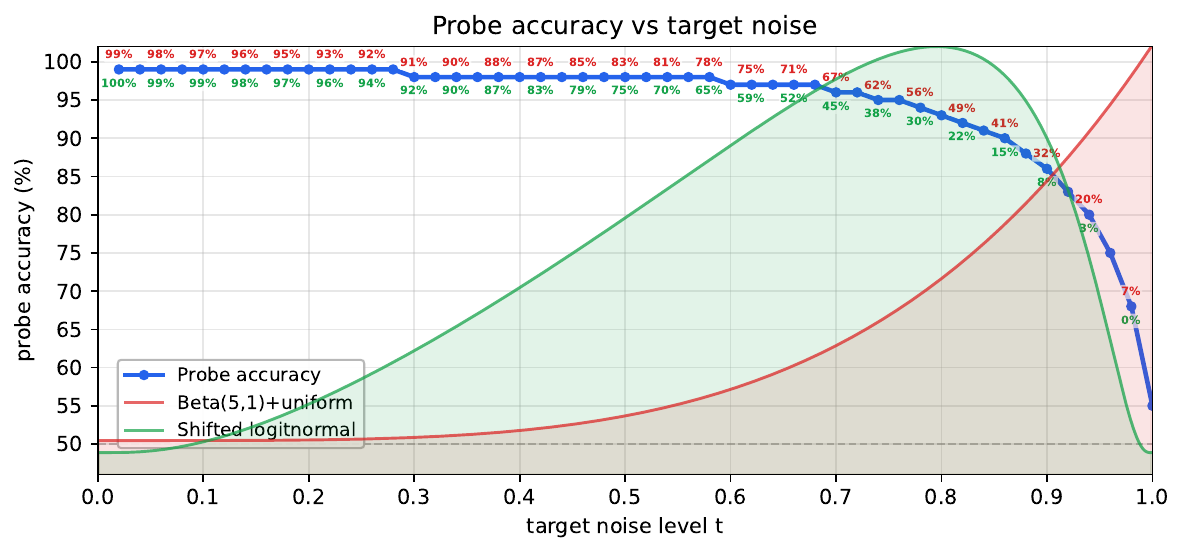}
    \caption{
        Reference-shortcut probe (\S\ref{sec:exp_probe}). Blue: probe binary-classification accuracy on the held-out set, measured at 50 evenly spaced target noise levels $t$; the gray dashed line marks chance ($50\%$). Light curves show two training-time timestep distributions on the same $t$ axis---our Beta+Uniform mixture (red) and a logit-normal distribution centered around $t = 0.8$ (green)---each independently rescaled to a fixed peak height. The small numbers above (red) and below (green) each blue marker give the fraction of training samples drawn at or above that noise level under each distribution, i.e.\ $\mathbb{P}[T \geq t]$.
    }
    \label{fig:probe_accuracy}
\end{figure}

Accuracy stays at $\geq 98\%$ for $t \leq 0.58$, remains $\geq 90\%$ through $t \approx 0.86$, and is still $75\%$ at $t = 0.96$.
Only as $t$ approaches $1$ does it collapse to chance.
Standard logit-normal distributions and their mildly shifted variants place the majority of training mass squarely in this high-accuracy regime, making the shortcut 
the easiest strategy for reducing the training loss.

At inference, generation starts from $t = 1$ (pure noise) and iteratively denoises.
The initial denoising steps, which establish the global structure of the output (including which speaker speaks where), occur at high $t$, where the shortcut is unavailable.
The model, however, has not learned to use text either.
The result is a model that fails to bind references to text, producing outputs where speakers are confused, ignored, or applied inconsistently.

\subsection{High-Noise-Biased Timestep Distribution}
\label{sec:recipe}

A na\"ive remedy is to re-center the logit-normal further toward $t = 1$.
The probe suggests this is insufficient.
The regime where the shortcut is solvable is wide enough that any reasonable shift still leaves the bulk of training mass inside it.
Pushing further into a narrow high-noise band fixes the shortcut but starves the model of the lower-noise signal it still needs for fine-detail synthesis.
We therefore replace the logit-normal with a mixture:
\begin{equation}
    p(t) = (1 - \lambda)\, \mathrm{Beta}(t;\, \alpha, 1) + \lambda\, \mathrm{Uniform}(t;\, \epsilon, 1).
    \label{eq:beta_uniform}
\end{equation}
The $\mathrm{Beta}(\alpha, 1)$ component, with density $\alpha t^{\alpha - 1}$, pushes substantial mass into the high-noise tail where the shortcut is least available; the Uniform component retains coverage across the rest of the range.

\subsection{Multi-Reference Dataset}
\label{sec:dataset}

We construct a multi-reference training dataset in which each example consists of a target clip, $K$ reference clips, and a caption describing how the full audio scene of the target is related to the references.
Each target is a multi-speaker conversational audio clip containing both dialog and non-speech sounds, while each reference is a distinct \emph{single-speaker} clip. 
This structure enables supervision in which a single natural-language description specifies how multiple reference speakers are realized within a shared conversational scene.
To construct this dataset, we employ a multi-stage pipeline comprising the following components:

\textbf{Matching references to target speakers.} 
For each audio clip, we extract speaker embeddings for each segment using a diarization pipeline. We then use embedding similarity to search for a separate reference clip in which each speaker appears. If a reference clip is found for every speaker, the original clip is defined as a target clip.
This process ensures alignment between the speakers in the target clip and its references.

\textbf{Intermediate target caption.} 
We first caption the target clip independently, without conditioning on the references. 
This intermediate caption provides a reliable description of the scene and reduces hallucinations compared to directly generating a reference-conditioned caption. 
It serves as a grounding signal for the final caption generation step.

\textbf{Multi-reference caption generation.}
Finally, we generate a caption for the target conversational scene conditioned on the reference clips.
The caption describes both the dialog and non-speech sounds, while grounding speaker identities in the references.
The captioning prompt is constructed by combining the reference clips, the intermediate target caption, and speaker-level timestamps obtained from the diarization process.
The intermediate caption provides a grounded description of the scene, reducing hallucinations, while the timestamps, aligned with the corresponding reference speakers, provide explicit temporal supervision over speaker turns.
Together, these signals improve speaker attribution and help the captioner produce captions that faithfully describe the full conversational scene.

\section{Experiments}
\label{sec:experiments}

\subsection{Experimental Setup}
\label{sec:setup}

\paragraph{Audio-only backbone.}
We adapt LTX-2~\cite{hacohen2026ltx2}, a dual-stream audio-video diffusion transformer with separate audio and video streams coupled by bidirectional cross-modal attention.
We use the LTX-2.3 release, retain only the audio stream, and remove all video-to-audio cross-attention layers, yielding a standalone audio-only diffusion transformer.
The variational autoencoder (VAE), text encoder, and prompt-embedding adapter of LTX-2 are reused without modification; references are encoded by the same VAE as the target.
The model supports audio durations of up to 20\,s.
\newpage
\paragraph{Training.}
We fine-tune the backbone to consume up to $K_\text{max} = 3$ speaker references per training example.
Reference clips can be of any length up to the model's 20\,s maximum, with only the target noised during training (\S\ref{sec:conditioning}).
Training proceeds for 20{,}000 steps at a global batch size of 128 on $16 \times$ NVIDIA GB200 GPUs, taking approximately 24 hours.
We optimize with AdamW ($\beta_1{=}0.9$, $\beta_2{=}0.95$, $\epsilon{=}10^{-8}$, weight decay $0.01$) at a peak learning rate of $1\!\times\!10^{-4}$, reached via a 1{,}000-step linear warmup and held constant thereafter.
We maintain an exponential moving average (EMA) of the parameters with decay $0.9999$.
The training objective is the rectified-flow velocity prediction~\cite{liu2023flow,lipman2023flow}.
Evaluations are performed on the EMA weights.

\paragraph{Auxiliary training-time augmentations.}
Two augmentations further tighten reference-to-text binding on top of the timestep distribution.
\emph{Adversarial reference injection} attacks the shortcut from a complementary angle.
At training time we fill empty reference slots with extra ``distractor'' references not mentioned in the prompt.
The only way to satisfy the prompt is then to bind by text rather than copy from whatever happens to be in the sequence.
No new loss term is introduced; the standard rectified-flow objective is unchanged, only the input.
\emph{Slot-shuffle augmentation} reduces positional bias on the reference slots.
At each step we permute the order of the references in the self-attention input sequence and rewrite the prompt's \texttt{reference $k$} tokens to match.
The model thus cannot rely on, e.g., ``slot~1 is always the first speaker''.
Our default uses both adversarial reference injection and slot-shuffle, with shuffle introduced as a curriculum: no shuffle for the first 10{,}000 steps (so the model first learns the basic mapping under a fixed slot order), shuffle thereafter.
We ablate both in Appendix~\ref{sec:exp_training_recipe}.

\paragraph{Evaluation set: \textsc{CoVoMix2-Dialogue-20s}.}
We evaluate on the public CoVoMix2 dialog test set~\cite{zhang2025covomix2}, which pairs 1{,}000 DailyDialog~\cite{li2017dailydialog} two-speaker transcripts with reference clips drawn from LibriSpeech \texttt{test-clean}~\cite{panayotov2015librispeech}.
We restrict to the 291 dialogs whose target fits the model's 20\,s budget, and call this subset \textsc{CoVoMix2-Dialogue-20s}.
The retained samples preserve the same speaker-gender mix and reference-similarity distribution as the full test set.
For every dialog, the two LibriSpeech prompt clips serve as $\mathbf{r}_1$ and $\mathbf{r}_2$, and the DailyDialog transcript is rendered into our standard \texttt{reference 1}/\texttt{reference 2} prompt format.

\paragraph{Reference robustness set: \textsc{CoVoMix2-Dialogue-WildRef}.}
\textsc{CoVoMix2-Dialogue-WildRef} probes realistic conditions beyond studio-clean LibriSpeech.
We sample 50 dialogs from \textsc{CoVoMix2-Dialogue-20s} and re-pair them with 30 in-the-wild English reference clips (crowd noise, background music, street ambience, wind, cartoon voices, and similar).
This yields 100 examples, with each wild clip used at least three times.
Holding the dialogs fixed isolates the effect of the reference distribution.

\paragraph{Metrics.}
We report seven metrics.
Three are standard: \textbf{WER} (Whisper-large-v3~\cite{radford2023whisper}), \textbf{UTMOS}~\cite{saeki2022utmos}, and \textbf{SQUIM}~\cite{kumar2023squim}.
The remaining four are multi-speaker variants central to our analysis.
\textbf{cpWER}~\cite{watanabe2020chime6} is a speaker-aware WER, computed post-diarization with best-permutation alignment so that attribution errors count.
\textbf{SIM-O} and \textbf{cpSIM} are cosine similarities between WavLM-ECAPA~\cite{chen2022wavlm} speaker embeddings, with cpSIM the strict per-speaker variant most sensitive to reference-speaker binding.
\textbf{ACC} is the fraction of words whose generated speaker (MMS forced alignment~\cite{pratap2024mms} + per-segment WavLM-ECAPA assignment) matches the prompt label.
Diarization uses pyannote.audio~\cite{bredin2023pyannote}.

\subsection{Comparison with Multi-Speaker Dialog Baselines}
\label{sec:exp_baselines}

We compare \method{} against current multi-speaker / dialog TTS systems on \textsc{CoVoMix2-Dialogue-20s}: MOSS-TTSD~\cite{zhang2026mossttsd}, VibeVoice-1.5B and VibeVoice-7B~\cite{peng2025vibevoice}, ZipVoice-Dialog~\cite{zhu2025zipvoicedialog}, and Dia (Nari Labs)~\cite{narilabs2024dia}.
All baselines are run with their public default settings.
As shown in Table~\ref{tab:baselines}, \method{} obtains the best cpWER, cpSIM, and ACC (the binding-aware metrics), together with the best WER and best (tied) SIM-O.
The two naturalness estimators diverge on \method{}: SQUIM places it at $4.32$ (within $0.02$ of the leaders), while UTMOS reads $3.44$ versus $3.76$ for MOSS-TTSD.
We attribute the UTMOS gap to the LTX-2.3 backbone, which is trained on in-the-wild video soundtracks rather than studio speech and inherits an acoustic profile that UTMOS scores more conservatively than SQUIM does.

\begin{table}[t]
\centering
\scriptsize
\setlength{\tabcolsep}{4.5pt}
\caption{Comparison with multi-speaker dialog baselines on \textsc{CoVoMix2-Dialogue-20s}.}
\label{tab:baselines}
\begin{tabular}{l ccc cccc}
\toprule
System & cpWER\,$\downarrow$ & cpSIM\,$\uparrow$ & ACC\,$\uparrow$ & WER\,$\downarrow$ & SIM-O\,$\uparrow$ & UTMOS\,$\uparrow$ & SQUIM\,$\uparrow$ \\
\midrule
MOSS-TTSD~\cite{zhang2026mossttsd} & 0.232 & \underline{0.547} & \underline{0.855} & 0.109 & 0.443 & \textbf{3.76} & 4.28 \\
VibeVoice-7B~\cite{peng2025vibevoice} & 0.206 & 0.527 & 0.821 & 0.044 & \textbf{0.451} & \underline{3.58} & 4.28 \\
VibeVoice-1.5B~\cite{peng2025vibevoice} & 0.212 & 0.503 & 0.830 & 0.050 & 0.423 & 3.56 & 4.27 \\
ZipVoice-Dialog~\cite{zhu2025zipvoicedialog} & \underline{0.176} & 0.538 & 0.847 & \underline{0.032} & \underline{0.446} & 3.57 & \textbf{4.34} \\
Dia (Nari Labs)~\cite{narilabs2024dia} & 0.303 & 0.339 & 0.757 & 0.133 & 0.312 & 2.69 & 4.09 \\
\midrule
\rowcolor{gray!12} \method{} & \textbf{0.145} & \textbf{0.567} & \textbf{0.866} & \textbf{0.020} & \textbf{0.451} & 3.44 & \underline{4.32} \\
\bottomrule
\end{tabular}
\end{table}
\newpage
\paragraph{Robustness to in-the-wild references.}
Table~\ref{tab:wildref} repeats the comparison on \textsc{CoVoMix2-Dialogue-WildRef}, where the studio-clean LibriSpeech prompts are replaced by 30 in-the-wild reference clips (50 dialogs, see \S\ref{sec:setup}).
\method{} retains the best cpSIM, SIM-O, WER, and SQUIM, and remains close to the leader on cpWER and ACC, where MOSS-TTSD edges ahead.
On the wild references, every baseline's cpSIM drops by roughly 0.15 absolute, falling below 0.40, while \method{} stays above 0.42.
The smaller open-source baselines (VibeVoice, Dia) fall off under the harder reference distribution.

\begin{table}[t]
\centering
\scriptsize
\setlength{\tabcolsep}{4.5pt}
\caption{Comparison on \textsc{CoVoMix2-Dialogue-WildRef} (50 dialogs paired with 30 in-the-wild reference clips).}
\label{tab:wildref}
\begin{tabular}{l ccc cccc}
\toprule
System & cpWER\,$\downarrow$ & cpSIM\,$\uparrow$ & ACC\,$\uparrow$ & WER\,$\downarrow$ & SIM-O\,$\uparrow$ & UTMOS\,$\uparrow$ & SQUIM\,$\uparrow$ \\
\midrule
MOSS-TTSD~\cite{zhang2026mossttsd} & \textbf{0.156} & 0.390 & \textbf{0.844} & 0.059 & 0.295 & \textbf{3.45} & \underline{4.21} \\
VibeVoice-7B~\cite{peng2025vibevoice} & 0.172 & 0.386 & \underline{0.841} & 0.045 & \underline{0.317} & 2.56 & 2.91 \\
VibeVoice-1.5B~\cite{peng2025vibevoice} & 0.202 & 0.365 & 0.826 & 0.089 & 0.293 & 2.33 & 2.85 \\
ZipVoice-Dialog~\cite{zhu2025zipvoicedialog} & 0.173 & \underline{0.396} & 0.825 & \underline{0.038} & 0.315 & 3.20 & 4.19 \\
Dia (Nari Labs)~\cite{narilabs2024dia} & 0.272 & 0.278 & 0.752 & 0.086 & 0.256 & 2.45 & 3.92 \\
\midrule
\rowcolor{gray!12} \method{} & \underline{0.167} & \textbf{0.424} & 0.819 & \textbf{0.022} & \textbf{0.348} & \underline{3.30} & \textbf{4.28} \\
\bottomrule
\end{tabular}
\end{table}

\subsection{Human Evaluation}
\label{sec:ab_preference}

We ran a side-by-side A/B preference test against the four baselines from \S\ref{sec:exp_baselines}, on items drawn from both \textsc{CoVoMix2-Dialogue-20s} and \textsc{CoVoMix2-Dialogue-WildRef}; the protocol is in Appendix~\ref{sec:subjective_eval}.
\method{} is preferred over every baseline at conventional significance (Table~\ref{tab:ab_results}).

\begin{table}[tb]
\centering
\scriptsize
\setlength{\tabcolsep}{10pt}
\caption{Side-by-side A/B preference results on a mix of \textsc{CoVoMix2-Dialogue-20s} and \textsc{CoVoMix2-Dialogue-WildRef} items. \emph{\method{} preferred} is \method{}'s win rate among non-tie ratings. Significance is from a two-sided binomial test ($^{\ast}\,p{<}0.05$, $^{\ast\ast}\,p{<}0.01$, $^{\ast\ast\ast}\,p{<}0.001$).}
\label{tab:ab_results}
\begin{tabular}{l c}
\toprule
Opponent & \method{} preferred \\
\midrule
ZipVoice-Dialog~\cite{zhu2025zipvoicedialog} & 84.6\%$^{\ast\ast\ast}$ \\
Dia~\cite{narilabs2024dia} & 74.2\%$^{\ast\ast\ast}$ \\
VibeVoice-7B~\cite{peng2025vibevoice} & 68.3\%$^{\ast\ast}$ \\
MOSS-TTSD~\cite{zhang2026mossttsd} & 59.8\%$^{\ast}$ \\
\bottomrule
\end{tabular}
\end{table}

\subsection{Noise Schedule Ablation}
\label{sec:ablation_noise}

\begin{figure}[!htb]
\centering
\scriptsize
\setlength{\tabcolsep}{4.5pt}
\begin{tabular}{l ccc cccc}
\toprule
& \multicolumn{3}{c}{Binding} & \multicolumn{4}{c}{Standard} \\
\cmidrule(lr){2-4} \cmidrule(lr){5-8}
System & cpWER\,$\downarrow$ & cpSIM\,$\uparrow$ & ACC\,$\uparrow$ & WER\,$\downarrow$ & SIM-O\,$\uparrow$ & UTMOS\,$\uparrow$ & SQUIM\,$\uparrow$ \\
\midrule
\rowcolor{leftmostcol!10} LogitNormal $\mu{=}0.17,\sigma{=}0.75$ & 0.167 & 0.503 & 0.830 & \underline{0.019} & 0.402 & \underline{3.64} & 4.29 \\
\rowcolor{leftcol!10} LogitNormal $\mu{=}0.62,\sigma{=}0.75$ & 0.158 & 0.500 & 0.850 & \underline{0.019} & 0.402 & \textbf{3.65} & 4.30 \\
\rowcolor{rightcol!10} LogitNormal $\mu{=}0.77,\sigma{=}1.0$ & \underline{0.154} & \underline{0.549} & \underline{0.859} & \textbf{0.018} & \underline{0.438} & 3.54 & \underline{4.30} \\
\rowcolor{betacol!10} \method{} (Beta+Uniform) & \textbf{0.145} & \textbf{0.567} & \textbf{0.866} & 0.020 & \textbf{0.451} & 3.44 & \textbf{4.32} \\
\bottomrule
\end{tabular}

\vspace{6pt}

\includegraphics[width=0.92\linewidth]{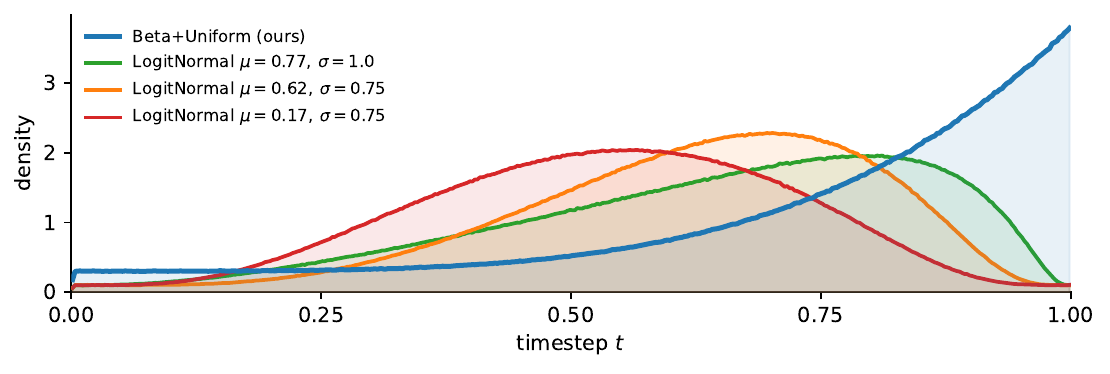}

\caption{Noise schedule ablation on \textsc{CoVoMix2-Dialogue-20s}. Top: results table, with row tints matching the curves below. Bottom: density of each sampler distribution; all variants share a 10\% uniform mix.}
\label{fig:noise_schedule_ablation}
\end{figure}

The probe illustrates that the shortcut is available; this experiment tests whether closing it at training time is what actually unlocks binding.
We compare our Beta+Uniform mixture (\S\ref{sec:recipe}) against three logit-normal distributions whose mass spans the range used in standard flow-matching practice, while keeping all other training settings fixed (\S\ref{sec:setup}).
The first row ($\mu{=}0.17, \sigma{=}0.75$) is closest to typical audio FM training; the remaining two progressively shift their mass toward higher noise.   Figure~\ref{fig:noise_schedule_ablation} shows the density of each timestep distribution (bottom) and the corresponding metrics (top).
All three binding-aware metrics (cpWER, cpSIM, ACC) improve monotonically as the schedule shifts toward higher noise.
\method{} leads on every binding metric while remaining competitive on the general audio-quality column.

\subsection{Qualitative Capabilities Beyond Per-Utterance Dialog}
\label{sec:qualitative}

The quantitative comparisons in \S\ref{sec:exp_baselines} hold the input format fixed to two-speaker dialog transcripts, the regime baselines were built for.
Our text-prompt format also enables generation modes that are difficult for per-utterance systems: \emph{overlapping speech} (unison readings, talk-overs, brief interjections), \emph{ambient and non-speech audio} generated jointly with the dialog, \emph{spontaneous paralinguistic events} (laughter, sighs, gasps, breaths) attributed in-line to named speakers, and \emph{multiple references per speaker} bound to a single voice for richer acoustic characterization.
We refer the reader to the project page for audio examples and to Appendix~\ref{app:spectrograms} for spectrograms of selected scenarios.

\section{Discussion}
\label{sec:discussion}

Contemporary multi-speaker systems typically route the binding problem through structured supervision: per-turn tags, multi-stream transcripts, identity encoders, or spatial layouts.
Our results show none of this is necessary.
A standard flow-matching transformer with concatenated reference latents and a single learned slot vector per index suffices, provided the timestep distribution is chosen to defeat the reference shortcut.
This minimalism is what unlocks the qualitative regimes in \S\ref{sec:qualitative}; no system with a structured interface can attempt them.

The shortcut argument itself is not specific to audio: it requires only (i) one or more clean references concatenated with a noised target, (ii) the noised target retaining enough information for similarity-based selection at moderate noise, and (iii) training mass concentrated outside the high-noise tail.
These conditions are satisfied by any reference-conditioned flow-matching model.
We expect the diagnosis and the high-noise-biased fix to transfer to image and video reference conditioning, though direct cross-modality verification is future work.

\section{Limitations}
\label{sec:limitations}

We inherit two practical limits from the backbone: a 20\,s generation cap and $K_\text{max} = 3$ reference speakers, beyond which the self-attention sequence grows linearly with $K$.
Both are softer than they appear: in LTX-2, audio tokens are roughly an order of magnitude fewer than video tokens, so an audio-only configuration has substantial headroom, enough to extend duration with modest fine-tuning and to fit additional references (or to add a learned reference-side compressor) without redesigning the model.
A third limit is inherent to the flow-matching paradigm: generation duration must be set before sampling, so FM systems resort to either heuristics or user input.



\clearpage
\bibliographystyle{plainnat}
\bibliography{references}

\clearpage
\appendix
\section{Supplementary}
\label{sec:supplementary}

\subsection{Reference Shortcut Probe Setup}
\label{sec:exp_probe}

The probe is trained on a binary classification task: given the noised target $\mathbf{z}_t$ and two candidate references (one drawn from the same sample as the target, the other from a different sample), predict which of the two matches the target.
We take the first 8 of the 48 transformer blocks of our backbone (discarding the rest) and attach a small two-layer MLP classification head on the pooled output.
Inputs are routed exactly as in the full model.
The target caption enters through the text cross-attention.
The noised target $\mathbf{z}_t$ and the two references form the self-attention sequence, with the references shuffled per example.
The timestep $t$ is supplied through the backbone's standard conditioning.
The head outputs a single logit predicting which of the two reference slots belongs to the target.
We train the probe for 10{,}000 steps at batch size 128, sampling $t$ uniformly in $[0, 1]$, and evaluate accuracy on $256$ held-out examples at each of $50$ evenly spaced timesteps in $[0, 1]$.

\subsection{Qualitative Spectrograms}
\label{app:spectrograms}

\begin{figure}[h]
\centering
\includegraphics[width=\linewidth]{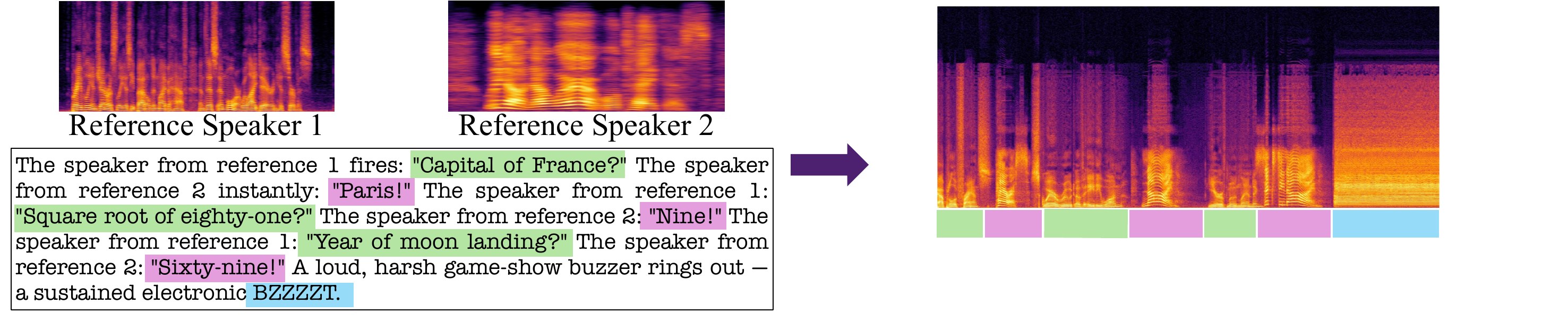}
\caption{Rapid-fire quiz closed by a non-speech buzzer.}
\label{fig:spectro_quiz}
\end{figure}

\begin{figure}[h]
\centering
\includegraphics[width=\linewidth]{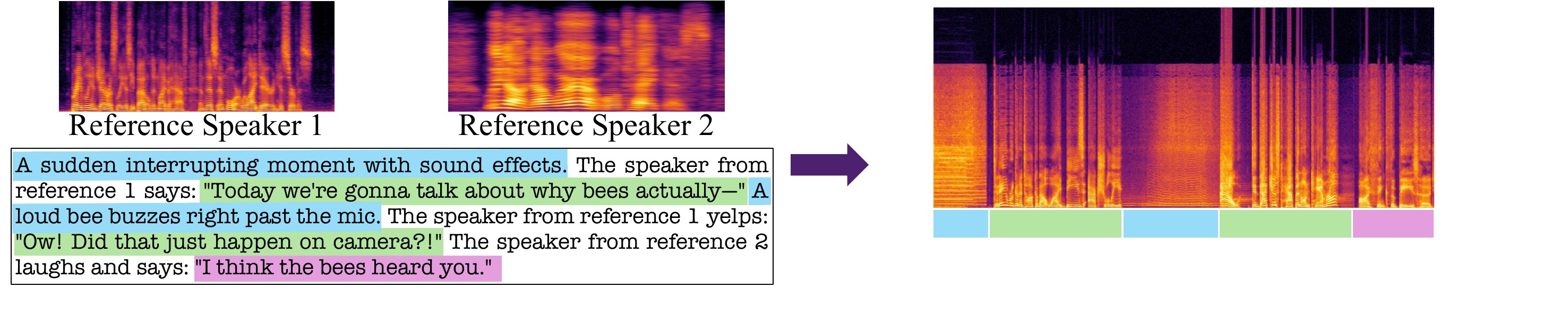}
\caption{Scene-level prompt with a mid-utterance ambient interruption and paralinguistic events.}
\label{fig:spectro_bees}
\end{figure}

\subsection{Reference Slot Encoding}
\label{sec:exp_slot_encoding}

We compare three slot-encoding choices.
The first is our additive slot embedding (Eq.~\ref{eq:abs_emb}).
The second is a RoPE-based alternative that augments the rotary positional encoding~\cite{su2021roformer} with an additional dimension.
The new dimension takes value $k$ for tokens belonging to reference $\mathbf{r}_k$ and value $0$ for tokens of the target $\mathbf{z}_t$.
The original temporal RoPE continues to index time within each segment, while the new dimension distinguishes the target from each reference:
\begin{equation}
    \text{RoPE}(\mathbf{x}) = \text{RoPE}_\text{temporal}(\tau) \otimes \text{RoPE}_\text{ref}(k),
    \label{eq:rope}
\end{equation}
where $\otimes$ denotes concatenation along the per-dimension rotations, $\tau$ is the within-segment time of token $\mathbf{x}$, and $k$ is its slot index.
The third is a \emph{no-positional} baseline that concatenates references in the self-attention sequence without any additional encoding to distinguish them from the target or from each other.

Adding a new RoPE dimension changes the positional statistics seen by the pretrained backbone.
For a fair comparison, we precede RoPE fine-tuning with a 5{,}000-step warmup on our general-audio corpus, with no references but the extra RoPE dimension already in place.
This lets the backbone adapt to the modified positional encoding before reference conditioning is introduced.
The additive and no-positional variants require no such adaptation and start fine-tuning directly from the backbone.

Table~\ref{tab:slot_encoding} reports results on \textsc{CoVoMix2-Dialogue-20s}.
The additive embedding leads on every binding-aware metric (cpWER, cpSIM, ACC); the RoPE variant follows closely.
The no-positional baseline collapses on binding: ACC falls to $0.513$ (essentially chance for two speakers) and cpSIM drops by $0.16$.
WER and naturalness are largely unchanged, since the model can still produce coherent speech when freed from the binding constraint.
This indicates that some explicit slot signal is necessary, but binding does not hinge on its precise form.

\begin{table}[t]
\centering
\scriptsize
\setlength{\tabcolsep}{5pt}
\caption{Reference slot encoding ablation on \textsc{CoVoMix2-Dialogue-20s}. The RoPE variant uses an additional 5{,}000-step warmup on general audio with the extra RoPE dimension in place.}
\label{tab:slot_encoding}
\begin{tabular}{l ccc cccc}
\toprule
System & cpWER\,$\downarrow$ & cpSIM\,$\uparrow$ & ACC\,$\uparrow$ & WER\,$\downarrow$ & SIM-O\,$\uparrow$ & UTMOS\,$\uparrow$ & SQUIM\,$\uparrow$ \\
\midrule
\method{} (no-pos) & 0.232 & 0.403 & 0.513 & \textbf{0.018} & 0.333 & \textbf{3.60} & \underline{4.28} \\
\method{} (RoPE) & \underline{0.181} & \underline{0.547} & \underline{0.835} & \underline{0.020} & \underline{0.449} & \underline{3.58} & \textbf{4.32} \\
\rowcolor{gray!12} \method{} (additive) & \textbf{0.145} & \textbf{0.567} & \textbf{0.866} & \underline{0.020} & \textbf{0.451} & 3.44 & \textbf{4.32} \\
\bottomrule
\end{tabular}
\end{table}

\subsection{Training Recipe Ablations}
\label{sec:exp_training_recipe}

Table~\ref{tab:training_recipe} ablates the two auxiliary augmentations described in \S\ref{sec:setup} (adversarial reference injection and slot-shuffle augmentation) on \textsc{CoVoMix2-Dialogue-20s}.
Removing adversarial references (\textit{no adversarial}) drops cpSIM by $\sim$0.10 and SIM-O by $\sim$0.08, with WER and naturalness essentially unchanged, confirming that the augmentation specifically tightens reference-speaker fidelity rather than generation quality.
Reference-shuffle behaviour is more nuanced: \textit{always-shuffle} from step~0 collapses (ACC $0.50$, near chance), because the model never gets a chance to anchor a stable reference-to-slot mapping before the augmentation starts moving slots around.
\textit{No-shuffle} at all is competitive on cpWER and ACC, but our curriculum retains a clear lead on the strict speaker-fidelity metrics (cpSIM, SIM-O, SQUIM) where positional bias would surface in adversarial / wild-reference conditions.

\begin{table}[t]
\centering
\scriptsize
\setlength{\tabcolsep}{5pt}
\caption{Training-recipe ablation on \textsc{CoVoMix2-Dialogue-20s}. \emph{no adversarial}: same recipe as ours but without random distractor references. \emph{always-shuffle}: reference-slot shuffle augmentation from step 0. \emph{no-shuffle}: never shuffle. Our default uses both adversarial distractors and a shuffle curriculum (no shuffle for the first 10\,k steps, shuffle thereafter).}
\label{tab:training_recipe}
\begin{tabular}{l ccc cccc}
\toprule
System & cpWER\,$\downarrow$ & cpSIM\,$\uparrow$ & ACC\,$\uparrow$ & WER\,$\downarrow$ & SIM-O\,$\uparrow$ & UTMOS\,$\uparrow$ & SQUIM\,$\uparrow$ \\
\midrule
\method{} (no adversarial) & 0.157 & 0.467 & 0.859 & \textbf{0.018} & 0.368 & 3.52 & \underline{4.29} \\
\method{} (always-shuffle) & 0.232 & 0.402 & 0.502 & \underline{0.019} & 0.334 & \textbf{3.68} & 4.26 \\
\method{} (no-shuffle) & \textbf{0.131} & \underline{0.491} & \textbf{0.886} & \textbf{0.018} & \underline{0.380} & \textbf{3.68} & 4.28 \\
\rowcolor{gray!12} \method{} & \underline{0.145} & \textbf{0.567} & \underline{0.866} & 0.020 & \textbf{0.451} & \underline{3.44} & \textbf{4.32} \\
\bottomrule
\end{tabular}
\end{table}

\subsection{Subjective Evaluation}
\label{sec:subjective_eval}

Table~\ref{tab:ab_instructions} shows the instructions presented to
evaluators for the side-by-side preference test. Each evaluator is given
two reference recordings of the original speakers (natural, noisy
recordings) followed by two synthesized versions of a dialog between
them, denoted (a) and (b). For every comparison the evaluator selects
(a), (b), or ``About the same''. Within a session, our system is paired
against one randomly chosen competitor per question, with the (a)/(b)
order randomized per item; evaluators are blind to system identity.

\begin{table}[h]
  \centering
  \caption{Side-by-side A/B Preference Evaluation Instructions.}
  \label{tab:ab_instructions}
  \begin{tabular}{p{0.92\linewidth}}
    \toprule
    \textbf{Instruction} \\
    \midrule
    You'll hear recordings of two original speakers (natural recordings
    with background noise; ignore the noise). Then you'll hear 2
    synthesized versions of a dialog between them. Pick the version
    where the voices sound more like the original speakers and the
    conversation feels more natural. \\[4pt]
    \textbf{Which one is better?} \\
    \quad (a): Dialog (a) is better \\
    \quad (b): Dialog (b) is better \\
    \quad About the same: Cannot tell which is better \\
    \bottomrule
  \end{tabular}
\end{table}

\end{document}